\documentclass[11pt,a4paper]{article}

\usepackage{capt-of}
\usepackage{jinstpub}
\usepackage{lineno}
\usepackage[numbers,sort&compress]{natbib}
\usepackage{siunitx}
\usepackage{wrapfig}
\usepackage{xcolor}
\usepackage{xspace}

\definecolor{darkblue}{RGB}{42, 96, 153}

\newcommand{\degc}{~\si{\celsius}}
\newcommand{\gbps}{~\si[per-mode=symbol]{\giga\bit\per\second}\xspace}

\newcommand{\tbyps}{~\si[per-mode=symbol]{\tera\byte\per\second}\xspace}
\newcommand{\ghz}{~\si{\giga\hertz}\xspace}
\newcommand{\mps}{~\si[per-mode=symbol]{\meter\per\second}\xspace}
\newcommand{\mm}{~\si{\milli\meter}}
\newcommand{\mum}{~\si{\micro\meter}\xspace}
\newcommand{\watt}{~\si{\watt}\xspace}
\newcommand{\aunit}[1]{~\si{#1}}

\title{Technical challenges designing a prototype common readout board for LHCb future upgrades}
\author{Julien~Jiro~Langouët on behalf of the PCIe400 development collaboration}
\affiliation{Aix Marseille Univ, CNRS/IN2P3, CPPM, Marseille, France}
\emailAdd{langouet@cppm.in2p3.fr}


\abstract{The LHCb Upgrade~I implemented a triggerless data acquisition system, employing a versatile readout board performing data protocol conversion and first stage of event building. It is a key component of the chain as the same hardware is used for readout of all sub-detectors, fast command and LHC clock distribution to front-end electronics.
\\
LHCb Upgrade~II will increase the global throughput by five times. Also, the transition to 5D techniques, exemplified by the FastRICH detector, requires phase deterministic clock distribution to front-end electronics. The PCIe400 development with an output bandwidth of 400\gbps over PCIe Gen~5 or 400GbE, and up to 48~bidirectional links at 25\gbps, significantly outperforms current systems. Its FPGA embeds 4~million logic elements and 32\aunit{\giga\byte} of memory, potentially providing a twelvefold increase in processing capability compared to existing PCIe40 board developed for LHCb Upgrade~I. Latest available PLL with intrinsic jitter <100\aunit{\femto\second}~pk-pk in combination with logic to measure and control clock phases that should enable reaching ultimate precision at $\mathcal{O}(10)\aunit{\pico\second}$.
The design process went through meticulous studies on three major topics : Thermal dissipation, Power distribution and Signal Integrity. An early power consumption resulted with a total power dissipated up to 220\watt which is more than 3~times what the previous generation board had been designed for. The study compares solid and heat-pipe heat-sink base in Computational Fluid Dynamics simulations to define the parametric limits of an air cooled solution targeting our application.
\\
PCB thickness limits imposed by the PCIe form factor also puts high constraints over the stack-up choice. Indeed, power estimation showed power rails sinking up to 100\aunit{A}. Furthermore, 18 of the 22 power rails on-board are dedicated to the FPGA while power regulators can not be placed optimally all around the FPGA. A systematic verification of voltage drop in all power planes allowed to fit a solution matching tolerances on all FPGA power supplies. In the end, a controlled-depth milled edge connector PCB unlocked the solution to satisfy power distribution integrity.
\\
Finally, 112\gbps PAM4 striplines have 50\ghz bandwidth. Designing such striplines requires peculiar attention not to create impedance mismatch. S-parameter extraction using classic 2D simulator showed different results in terms of attenuation compared to new 3D simulators considering full geometry of vias. The counterpart of 3D simulators being the processing times. We study on how to reduce simulation time from 10 days to 10 hours. This work allowed to run comparison between various geometries focusing on breakout zones.
}

\begin{document}   


\maketitle

    \section{Module architecture and features}

The LHCb Upgrade I is designed to run at an instantaneous luminosity of $2\times10^{33}cm^{-2}s^{-1}$, achieved with a bunch-crossing rate of up to 40 MHz and a mean number of proton-proton interactions per bunch crossing of six, producing a data throughput of 5\tbyps. With the HL-LHC upgrade, the detector will operate at a maximum instantaneous luminosity of $1.5\times10^{34}cm^{-2}s^{-1}$, producing $\sim40$ proton-proton interactions per bunch crossing with a data throughput of 25\tbyps.~\cite{LHCbcollaboration:2886764}

The LHCb data acquisition system foreseen for the Upgrade~II is similar to the one developed for the Upgrade~I. The yield of event useful for physics is maximized by using a fully synchronous readout of each bunch crossing. This architecture is called triggerless because there is no more trigger decision sent to the front-end electronics. Instead, the data acquisition system consist of a single stage detector readout, followed by event-building on local area network and by real time reconstruction and selections of events. The system relies on a single custom made readout board, together with Commercial Off the Shelf (COTS) data-center hardware.~\cite{CERN-LHCC-2014-016}

The prototype readout board developed, called PCIe400, is at the interface between the front-end electronics on detector using radiation hardened protocols and back-end electronics using commercial data-center protocols. The same hardware, with different gatewares, is used for readout of all sub-detectors, fast command and LHC clock distribution to front-end electronics.

\begin{figure}[!tb]
    \centering
    
    \includegraphics[width=0.53\textwidth]{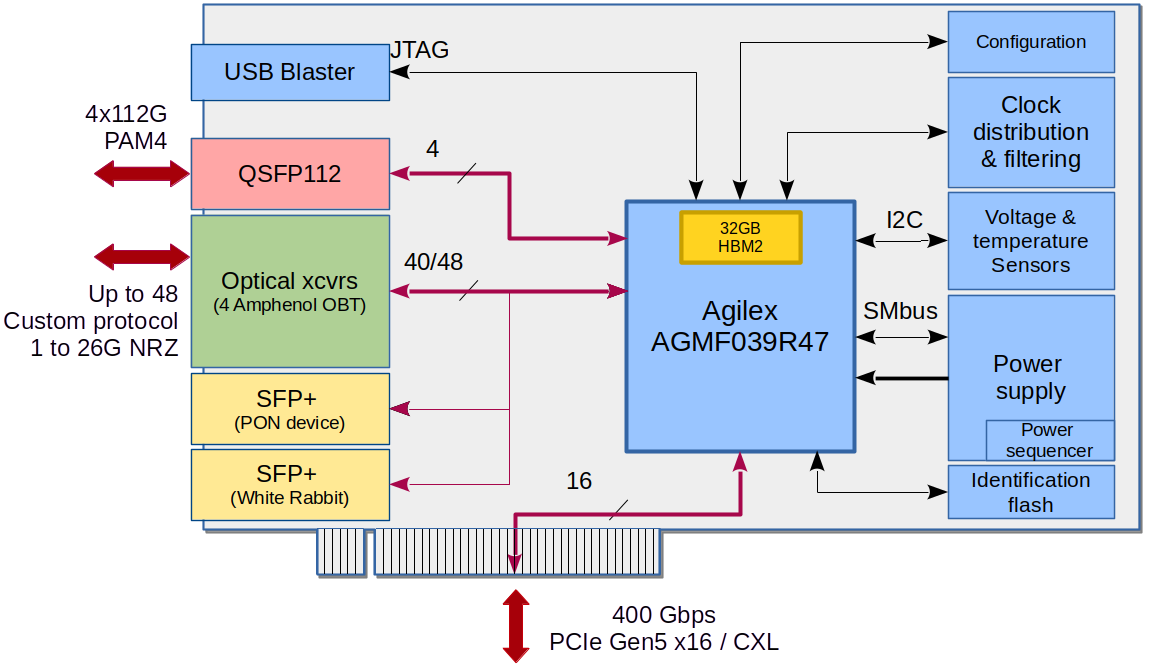} \hfill
    \raisebox{0.6\height}{\includegraphics[width=0.4\textwidth]{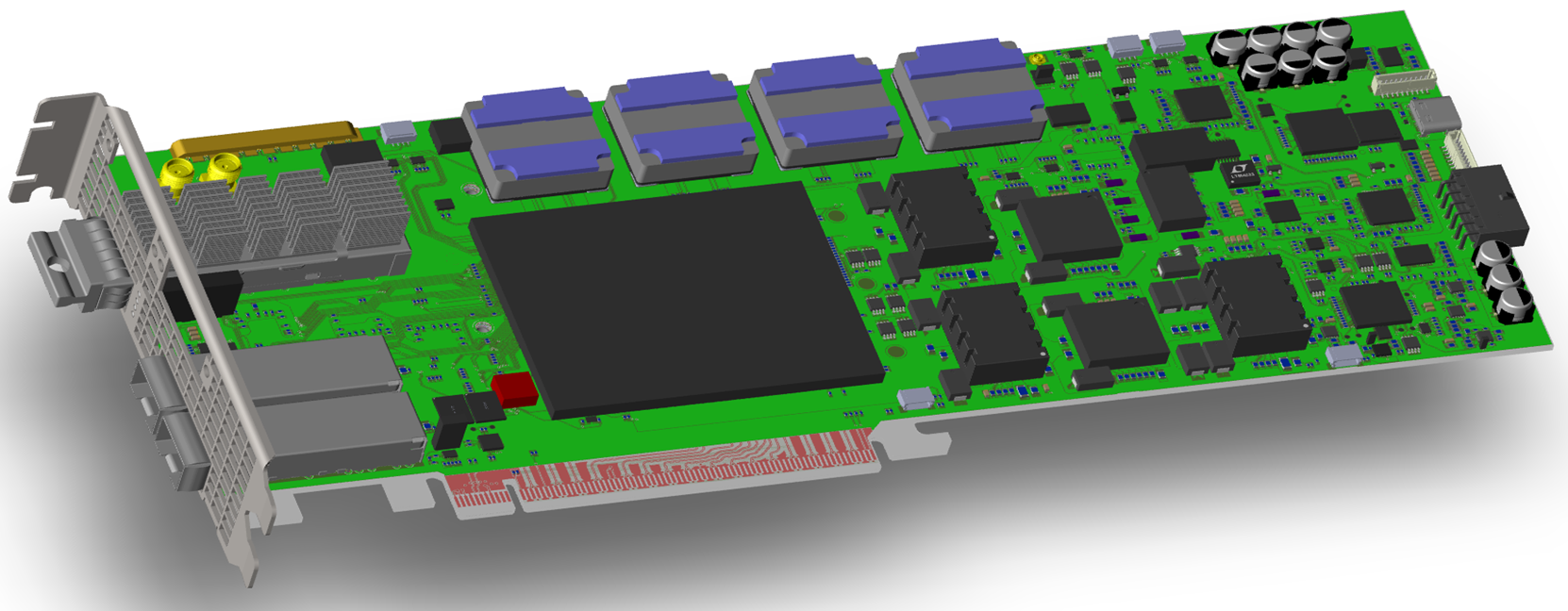}}
    \caption{PCIe400 Synoptic and 3D layout view.}
    \label{fig:pcie400_synoptic_3D}
\end{figure}

The board includes Altera's flagship FPGA, Agilex 7 M-series, with 4~million logic elements running at up to 1\ghz, and embeds 32\aunit{\giga\byte} of high bandwidth memory.~\cite{ALTERA-AGILEX} It features up to 48 bidirectional links for custom protocols between 1 and 26\gbps each.
It is connected to a PCIe Gen~5~x16 bus, giving it an effective bandwidth >~400\gbps with the host server CPU. It also implements an optional 400G~Ethernet bus over a QSFP112 form factor.
Finally, it manages various protocols associated with time distribution, such as TTC-PON\footnote{Timing, Trigger and Control Passive Optical Network} and White Rabbit.~\cite{9374437}~\cite{Serrano:1215571} A synoptic and a 3D layout view of the board is given in figure~\ref{fig:pcie400_synoptic_3D}. 
\\
Routing such board represents a challenge due to high density of components. In fact, there are \num{2500} components with \num{10000} connections on a 18 layers $268\times100\mm$ PCB. In particular, the FPGA ball grid array is \num{4500} pins over a $56\times66\mm$ package area which represents a 25\% increase in density compared to previous generation board using Altera's Arria 10 FPGA family thanks to a hexagonal ball grid array.

    \section{Thermal dissipation}


Using Altera Power and Thermal Calculator and scaled resource usage on new FPGA compared to current LHCb readout system, we define a typical\footnote{typical case defined as 12.5\% toggle rate, 640\aunit{MHz}, 60\% logic, 80\% RAM, 48~links @10\gbps} and worst case\footnote{worst case defined as 12.5\% toggle rate, 640\aunit{MHz}, 80\% logic, 100\% RAM, 40~links @25\gbps + 400GbE}.\cite{PTC} The dissipated power estimated is between 120 to 230\watt varying with the junction temperature due to leakage current, as show in figure~\ref{fig:fpga_power}. It is therefore important not to under-design the heat-sink, while an air cooling solution has been privileged for its simplicity on infrastructure. 

\noindent
\begin{minipage}{0.42\textwidth}
    \includegraphics[width=\textwidth]{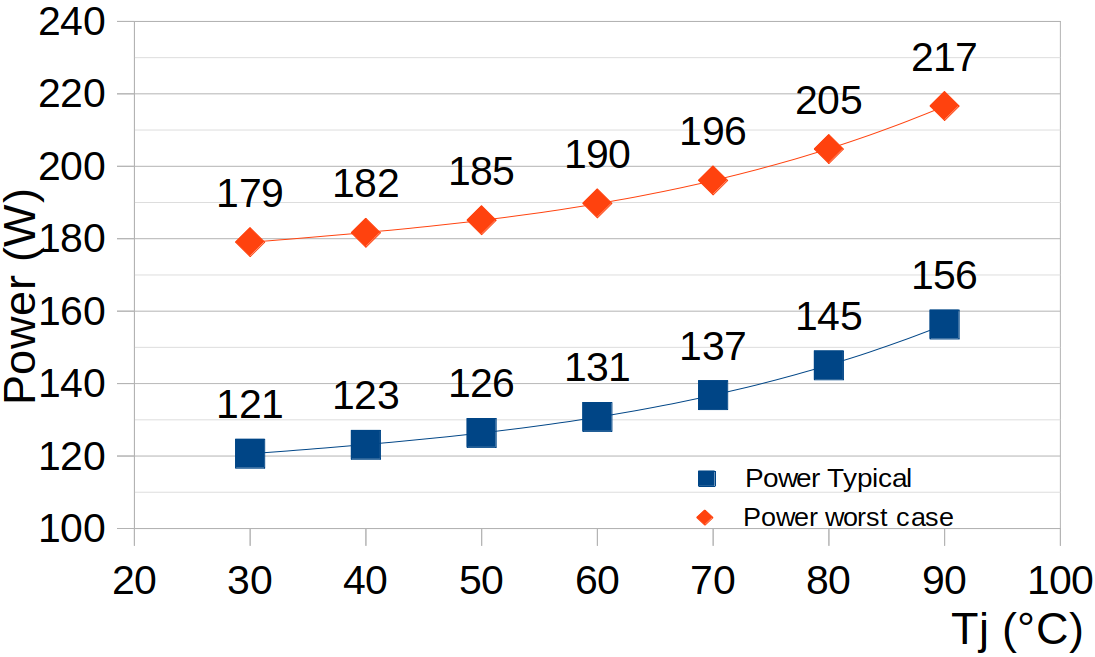}
    \captionof{figure}{FPGA power dissipation estimated for typical and worst case vs junction temperature.}
    \label{fig:fpga_power}
\end{minipage}
\hspace{0.02\textwidth} 
\begin{minipage}{0.56\textwidth}
    \includegraphics[width=\textwidth]{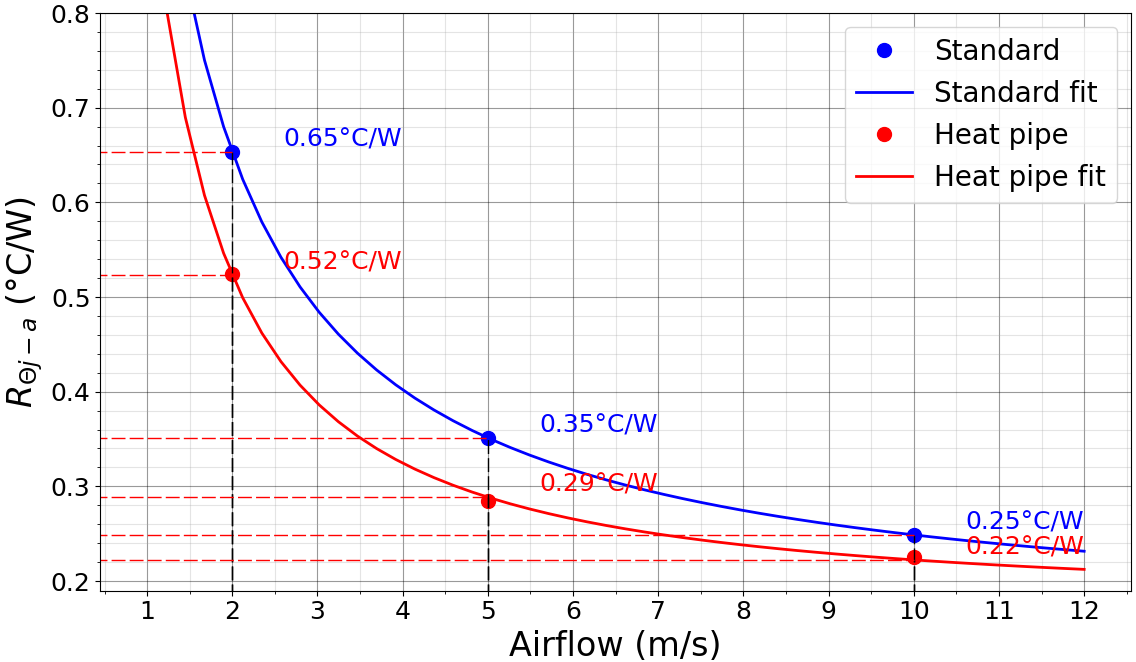}
    \captionof{figure}{FPGA heat-sink thermal resistance vs airflow comparing standard and heat-pipe heat-sink.}
    \label{fig:thermal_resistance}
\end{minipage}

Considering placement constraint on board and airflow direction along the length axis of the board, standard aluminum are compared to heat-pipe enhanced heat-sink in Computational Fluid Dynamic (CFD) simulation. As shown on figure~\ref{fig:thermal_resistance}, with ambient temperature of $38\degc$, airflow of $5\mps$, the overall thermal resistance of the heat-sink is lowered by 20\% thanks to reduced spreading resistance across the heat-sink base. In this condition, heat-sink can dissipate 135 on a standard and 165\watt on a heat pipe heat-sink, and maintain FPGA junction temperature at 85\degc. 

\begin{figure}[!tb]
    \centering 
    
    \includegraphics[width=0.8\textwidth]{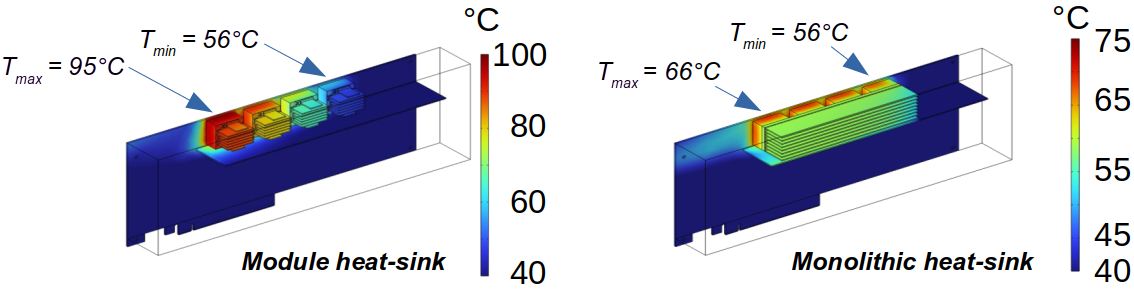}
    \caption{CFD heatmap of On-board opto-electronic transceiver comparing module and monolithic heat-sink}
    \label{fig:obt_cfd}
\end{figure}

On the side of the FPGA, the on-board opto-electronic transceiver (OBT) consume 30\watt over 4~modules. The placement constraint results in a heat cumulative effect which standard heat-sink provided by manufacturer can not handle. As shown in figure~\ref{fig:obt_cfd}, module heat-sink creates turbulence in between modules affecting airflow passing through. Instead a custom monolithic heat-sink reduce the average temperature by 20\%. 

CFD simulation conditions and results are described in table~\ref{tab:CFD_simulation}.

\begin{table}[!tb]
        \centering
          \caption{CFD simulation conditions and results, Ambient temperature $=38\degc$.}
        \begin{tabular}{|c|c|c|}
             \hline
             Condition                          & Typical       & Worst             \\ \hline
             Air flow                           & 5\mps          & 10\mps           \\ \hline
             FPGA power dissipation / $T_{j}$ 
                                                & 160\watt / 84\degc  & 220\watt / 87\degc      \\ \hline
             OBT power dissipation / $T_{\rm case}$ & 30\watt / 57\degc   & 30\watt / 52\degc       \\ \hline
             QSFP112                            & 13\watt / 58\degc   & 13\watt / 61\degc       \\ \hline
        \end{tabular}
        \label{tab:CFD_simulation}
    \end{table}

    \section{Power Integrity}

\begin{wrapfigure}{r}{0.5\textwidth}  
    \includegraphics[width=0.5\textwidth]{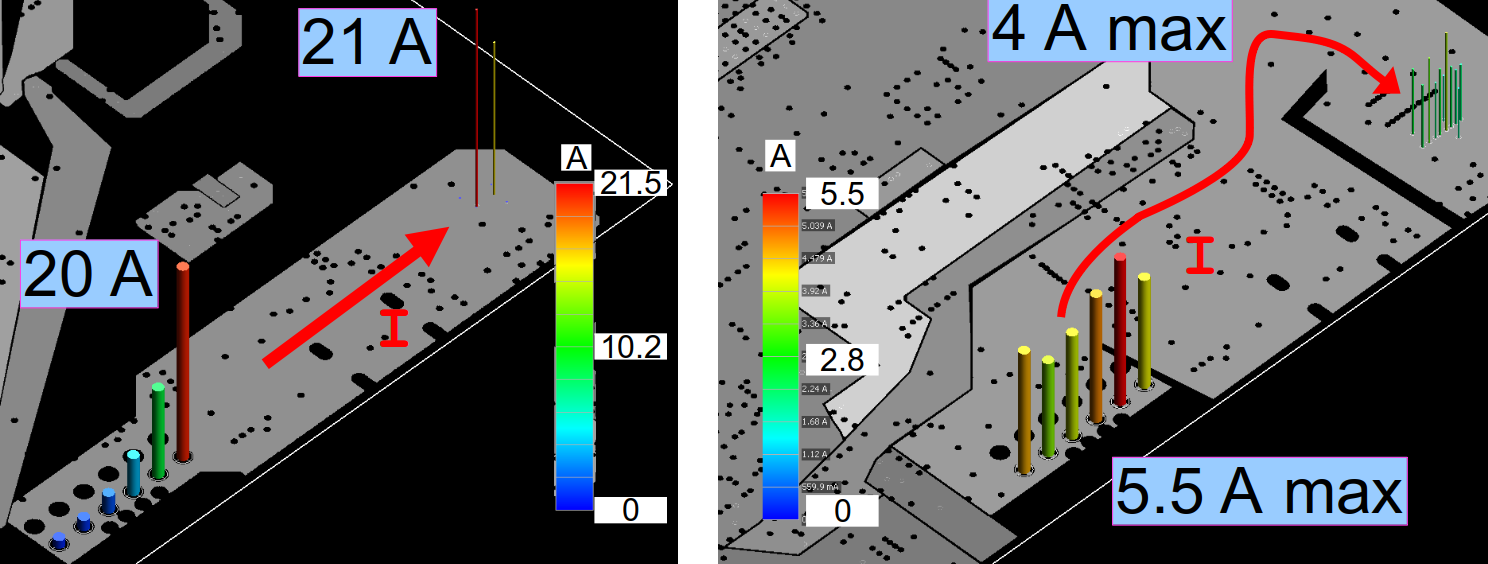}
    \caption{(left) original vs (right) optimized power plane design reducing maximum current in single via.}
    \label{fig:power_plane_opt}
\end{wrapfigure}

There are 22 power rails ranging from 0.8 to 5\aunit{V} with 0.5\% accuracy and current up to over 100\aunit{A}. A systematic voltage drop analysis post layout on each power rails at absolute maximum current rate is required to prevent design flaws affecting board reliability and performance. Points of attention are to minimize heat loss in power planes by equalizing current densities in planes and vias. One example of power plane optimization is illustrated in figure~\ref{fig:power_plane_opt}. In this case, increasing current path length and number of vias is required in order to reduce current handled by a single via. 

\begin{wraptable}{r}{0.45\textwidth}
    \caption{Thermal dissipation in PCB simulation.}
    \begin{tabular}{|c|c|c|}
        \hline
        Top \& L9 thickness &  70\mum	&  35\mum \\ \hline
        Total power loss &	11.2\watt &	18.7\watt \\ \hline
        $\Delta T$ PCB &	23\degc & 42\degc \\ \hline
    \end{tabular}
    \label{tab:thermal_dissipation_sim}
\end{wraptable}

Pre-design simulation study used the DK-DEV-AGI027RES design from Altera to demonstrate the effect of copper thickness on the PCB temperature increase due to heat dissipated in power planes.\cite{DK-DEV} The simulation takes into account the single dominating power rail with 200\aunit{A} at 0.8\aunit{V},  while other power supplies accumulated are negligible. 
We compare a stack-up using 2 copper planes of 35 or 70\mum. The results, given in table~\ref{tab:thermal_dissipation_sim}, show that the PCB temperature increase is divided by a factor 2 using 70\mum power planes. The simulation results oriented the design to use a controlled-depth milled edge connector PCB, in order to respect PCIe PCB thickness constraint of $1.57\mm\pm10\%$ while using 2 internal layers of 70\mum for main power rails.~\cite{pcie_spec_5}
    
    \section{Signal Integrity}

\begin{wrapfigure}{r}{0.5\textwidth}  
    \includegraphics[width=0.5\textwidth]{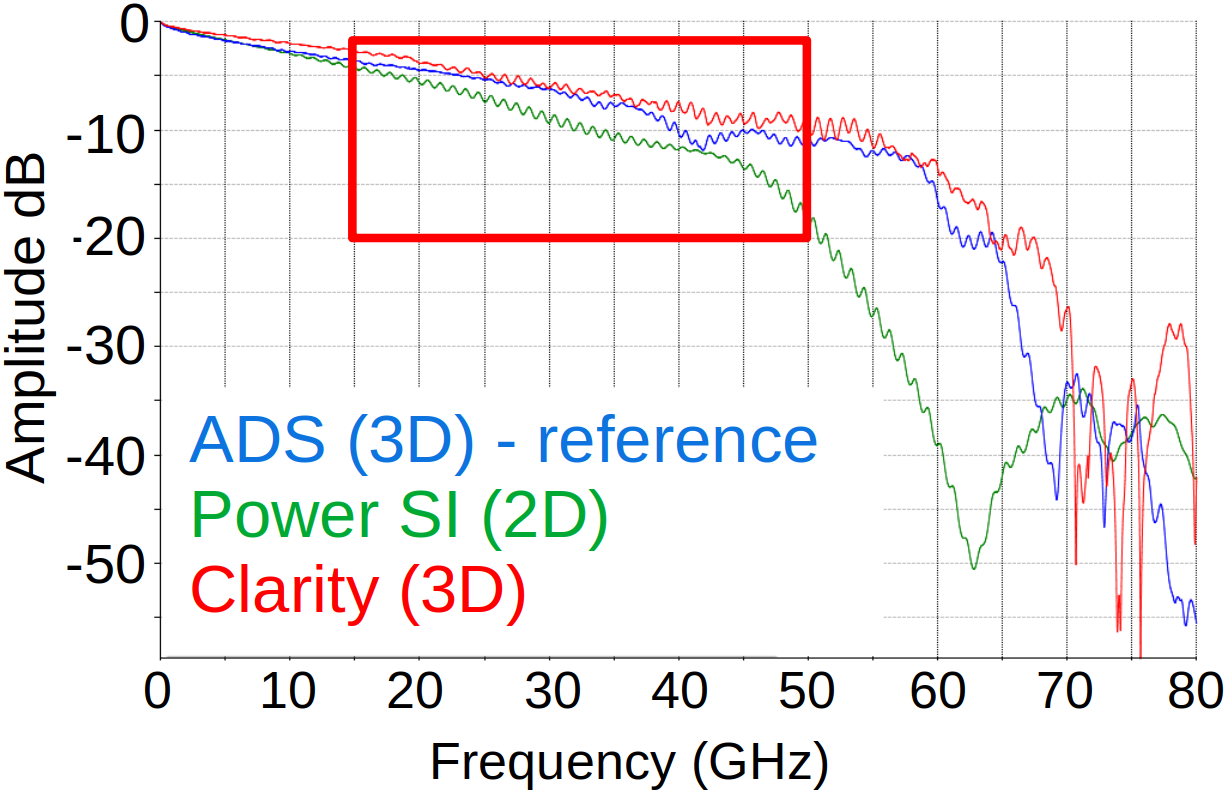}
    \caption{Comparison of insertion loss - SDD21 - of OSFP800 Tx~4 between simulation tool.}
    \label{fig:s-param_comp}
\end{wrapfigure}

There are 136 lanes on board ranging from 25\gbps NRZ\footnote{Non Return Zero signal modulation} to 112\gbps PAM4\footnote{Pulse Amplitude Modulation over 4 levels}. 
The serial links bandwidth is between 20 to 50\ghz.\cite{OIF-CEI} We compare Cadence Power SI and Cadence Carity 3D tools to a reference provided by Altera using Keysight ADS on a OSFP800 transmitter lane.~\cite{9528791}~\cite{DK-DEV} Looking at insertion loss on figure~\ref{fig:s-param_comp}, we can see a difference between 2D and 3D tools over 20\ghz. Using Cadence Clarity~3D comes with the need for high performance computational resources : its takes up to 6 hours to extract s-parameter of a $\sim10\aunit{cm}$ lane with 800 frequency points on a 64 cores @2.5\ghz machine. 

\begin{figure}[!b]  
    \centering
    
    \includegraphics[width=0.5\textwidth]{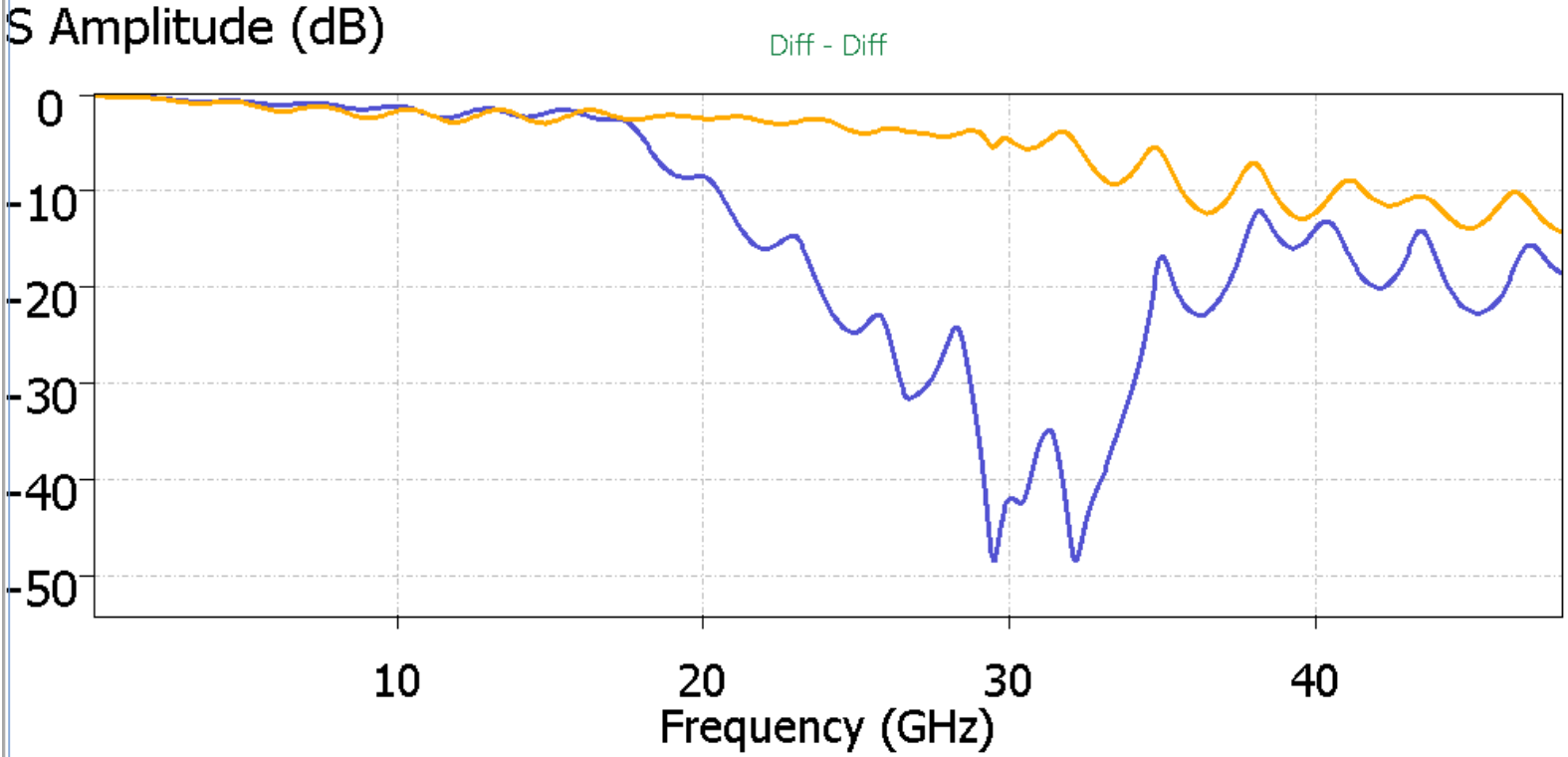} \hfill
    \includegraphics[width=0.35\textwidth]{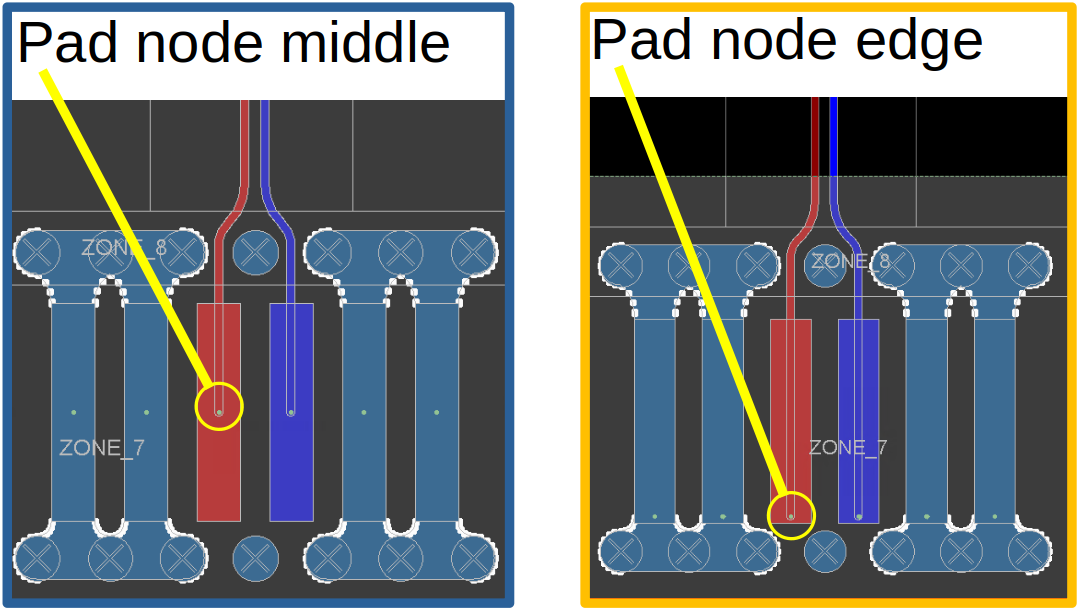}
    \caption{(left) Insertion loss - SDD21 - of PCIe Rx lane (\textcolor{darkblue}{blue}) with node pad middle, showing a stub resonance, (\textcolor{orange}{orange}) with node near pad end edge}
    \label{fig:sdd21_port_comp}
\end{figure}

Also, some features can be misleading using clarity~3D, as observed on figure~\ref{fig:sdd21_port_comp}, port placement within connector pads have a large effect on the insertion loss SDD21, while there is no consequence on layout level for fabrication. They are usually defined at the center of the pad. \\

%

3D s-parameter extraction can help verifying breakout regions to reduce impedance mismatch. In fact, breakout region near FPGA or opto-electronic transceiver are the most critical region because of layer transition. Following are two examples of optimizations. 

\paragraph{SMD connector breakout optimization}

\begin{wrapfigure}{r}{0.5\textwidth}  
    \centering
    
    \includegraphics[width=0.23\textwidth]{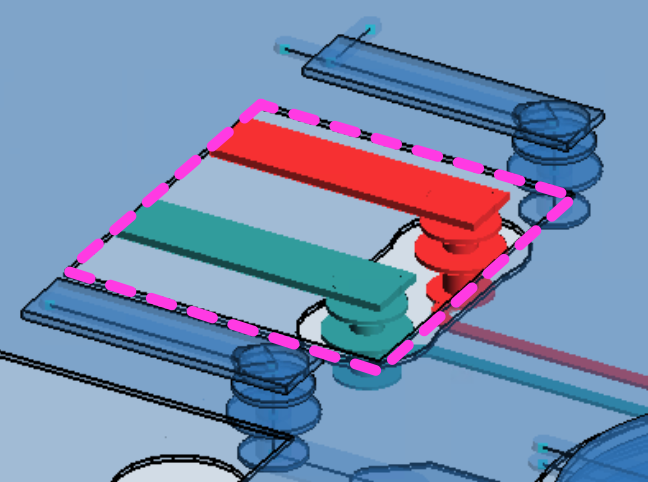}
    \includegraphics[width=0.5\textwidth]{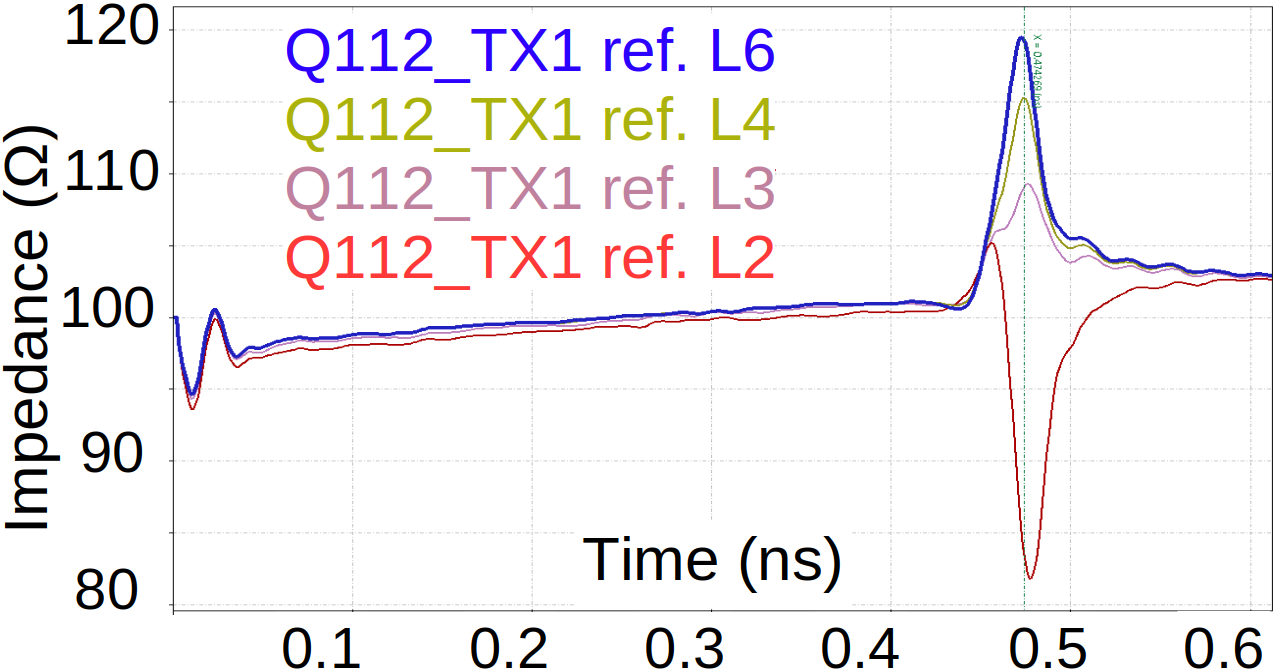}
    \caption{(top) SMD pad with cut-out to reference SMD to L3. (bottom) 112\gbps TDR response at $T_{rise}~=~7\aunit{ps}$, impedance mismatch vs distance to reference plane}
    \label{fig:smd_breakout}
\end{wrapfigure}

SMD pads of a QSFP connector creates a parasitic capacitance with planes underneath. It is inversely proportional to plane distance. By sweeping cut-outs across layers, as shown in figure~\ref{fig:smd_breakout}, best balance to reduce impedance mismatch can be found. The ideal solution may not be achieved depending on layer thickness but anti-pad size can help adjusting ac-coupling. In our case, with a cut-out down to layer 3, corresponding to 195\mum, the impedance remains within the $100\aunit{\Omega}\pm7\%$ target.\cite{ALTERA-HSSI-GUIDE}


\paragraph{FPGA breakout optimization}

\begin{wrapfigure}{r}{0.5\textwidth}  
    \includegraphics[width=0.5\textwidth]{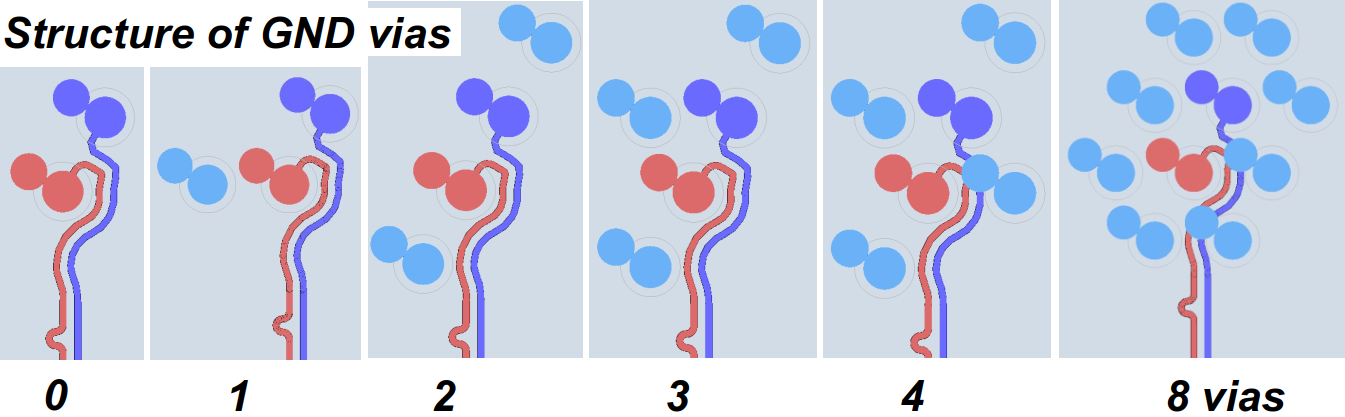}
     \includegraphics[width=0.5\textwidth]{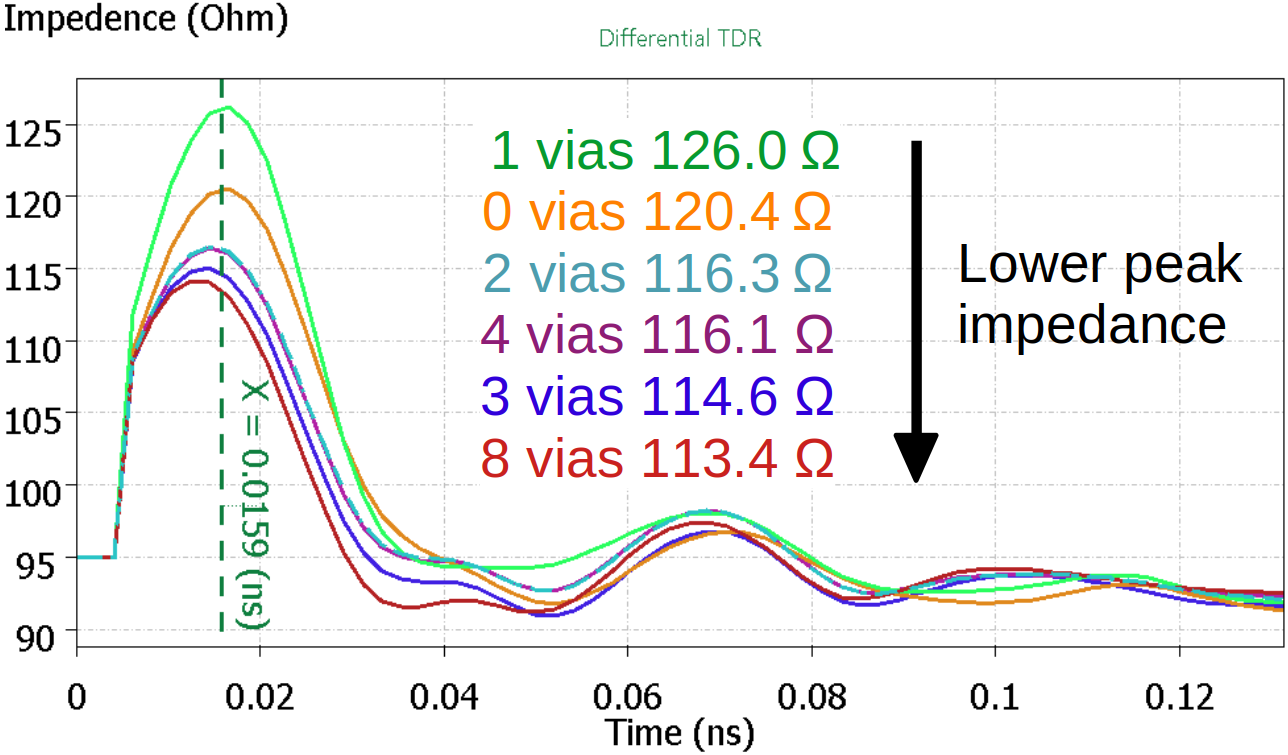}
    \caption{(top) view of ground vias structure to shield differential pairs (bottom) PCIe Tx FPGA breakout TDR response to $T_{rise}~=~15\aunit{ps}$, target 95\aunit{\Omega}}
    \label{fig:fpga_breakout}
\end{wrapfigure}

On layer transition, it is generally recommended to shield the differential pairs with ground vias to mimic a coaxial cable. However, in our case, we need to use staggered vias for PCIe Tx lanes to reach Layer 16 on secondary side of PCB. Because of high density in the region of FPGA breakout, the ideal case with 8 ground vias surrounding differential pair can not be achieved. By sweeping on different ground via structure varying the number of GND vias, time domain reflection simulation shown in figure~\ref{fig:fpga_breakout}, reveals that 3 vias symmetric structure is almost as well performing as the 8 vias structure with only 1\% relative difference. The use of reduced to 3 ground vias structure patterns for breakout ensured consistent signal integrity across several lanes. 

    \section{Summary}

The design process of the PCIe400 common readout board for LHCb future upgrade went through meticulous simulation work for thermal dissipation, power and signal integrity. CFD simulation resulted in designing custom heat-sink with heat-pipes capable of dissipating 165\watt with 5\mps airflow. A systematic power distribution network verification allowed to optimize power plane and via current to ensure reliability and performance. We demonstrated the interest of 3D tool for S-paramter extraction for over 20\ghz bandwidth signals to optimize critical breakout zone especially on unprecedentedly used 112\gbps PAM4 serial links.
    
\bibliographystyle{unsrt}
\bibliography{references}

\end{document}